**European coastal wetlands datasets and their use in decision-support tools for policy restoration objectives**

Bruna R.F. Oliveira[1*], Antonio Camacho[2], Anis Guelmami[3], Christoph Schroder[4], Nina Bègue [3], Martynas Bučas[5], Constantin Cazacu[6], Elisa Ciravegna[7], João Pedro Coelho[8], Relu Giuca Constantin[6], Samuel Hilaire[3], Marija Kataržytė[5], Daniel Morant[2], Antonio Picazo[2], Nico Polman [7], Marinka van Puijenbroek [7], Justine Raoult[7], Carlos Rochera[2], Michaël Ronse [3], Lisa Sella[9], Ana I. Sousa [8]; Tudor Racoviceanu[6], Francesca S. Rota[10], Diana Vaičiūtė [5], Ana I. Lillebø[8*],

[1]Centre for Environmental and Marine Studies, Department of Environment and Planning, University of Aveiro, Portugal

[2]Cavanilles Institute for Biodiversity and Evolutionary Biology, University of Valencia, E-46980 Paterna, Spain

[3]Tour du Valat Research Institute for the Conservation of Mediterranean Wetlands, France

[4]European Topic Centre for Spatial Analysis and Synthesis, University of Malaga, España

[5]Marine Research Institute, Klaipėda University, Universiteto ave. 17, 92294, Klaipėda, Lithuania

[6]Research Centre and Department of Systems Ecology and Sustainability, University of Bucharest

[7]Wageningen Social & Economic Research and Wageningen Marine Research, Netherlands

[8]ECOMARE, Centre for Environmental and Marine Studies, Department of Biology, University of Aveiro, Portugal

[9]CNR Research Institute on Sustainable Economic Growth, Torino, Italy

[10]Department of Economics and Statistics "Cognetti de Martiis", University of Turin, Italy

[*]Corresponding authors: bruna.oliveira@ua.pt; lillebo@ua.pt

## Abstract

Ecosystem restoration is a paramount policy priority for this decade, with ambitious global and European targets requiring unprecedented levels of data-driven implementation. Achieving effective and equitable restoration, particularly for coastal wetlands, hinges on spatially explicit socio-ecological information - maps that integrate habitats, ecosystem services, human activities, and pressures to guide prioritization, stakeholder negotiation, and adaptive management. This study, grounded in the RESTORE4Cs initiative, introduces an innovative multi-layered dataset that bridges science and policy for six emblematic European coastal wetlands: Ria de Aveiro (Portugal), Valencian Wetlands (Spain), Camargue (France), Southwest Dutch Delta (Netherlands), Curonian Lagoon (Lithuania), and the Danube Delta (Romania). The dataset consolidates ecological mapping (EUNIS 2021, 2022), human activity and pressure documentation (aligned with EU Habitats Directive, Water Framework Directive, Marine Strategy Framework Directive), comprehensive ecosystem services mapping (CICES v5.1), alongside robust participatory community and stakeholder data. Altogether, the database encapsulates 97 habitat records, 23,160 activity-pressure associations, and 1,668 ecosystem service records—enabling robust cross-regional analyses and direct integration into evidence-based decision support tools. By illustrating practical pathways for participatory engagement, trade-off negotiation, and cross-scale integration, this research equips scientists, policymakers, practitioners, and communities with the scientific foundation to propel the Nature Restoration Regulation and Biodiversity Strategy 2030 objectives, fortifying Europe's climate adaptation trajectory. The approach showcased signals a new era for restoration science – where spatially explicit, multi-actor data supports policy, mobilizes citizen stewardship, and accelerates the transformative ambitions of Europe's restoration decade.

## 1. Introduction

European coastal wetlands cover roughly 84,500 km², representing only one percent of EU27+UK wetland coverage, yet their contribution to climate mitigation and biodiversity far surpasses their extent [1]. Coastal wetlands, when in good ecological status, provide essential ecosystem services, including climate regulation (carbon pathways), shoreline stabilization, wave attenuation, biogeochemical regulation (nutrient cycling and water quality), and high taxonomic and functional diversity (complex trophic networks). The diversity of European coastal wetland habitats, as classified by EUNIS - a reference framework for EU habitat classification - encompasses subtidal seagrass beds and shallow marine sediments; intertidal flats, saltmarshes, and saline lagoons; and transitional waters, including estuaries and deltas. These habitats, shaped by regional biogeographical and hydrological regimes, span all biogeographical regions of Europe, reflecting varied climatic, geomorphological, and anthropogenic contexts.

IPBES (Intergovernmental Panel for Biodiversity and Ecosystem Services) notes that up to 50 percent of global coastal wetlands have been lost since 1900, with local losses exceeding 80 percent in some regions. Remaining wetlands face increasing vulnerability to: sea-level rise, coastal erosion, sediment deficits, droughts, altered hydrology, and land-use change pressures [2]. Wetlands face intensifying pressures from land reclamation, urban expansion, agricultural runoff, and hydrological and geomorphological modifications (e.g., activities that alter sediment supply, shorelines introduce pollutants, disrupt natural water flow). These anthropogenic activities and associated pressures compromise habitat status, decrease ecosystem service delivery, and increase vulnerability to climate extremes (persistent deviations in climate variables). Without habitat-specific pressure mapping, management efforts risk misallocating resources, as pressures like pollution, invasive species, and infrastructure development disproportionately affect some habitats [2]. This diagnosis has made the current decade the momentum for ecosystem, particularly coastal wetland, conservation and restoration. Relevant examples are UN Ocean Decade [3], UN Decade on Restoration [4], EU Biodiversity Strategy [5], and the EU Nature Restoration Regulation (NRR) [6], all with specific restoration targets for 2030. Wetland Restoration includes actions (active or passive) to return a wetland to a previous or improved ecological status, enhancing carbon storage, biodiversity, and ecosystem services [6]. Active restoration removes the pressure source followed by measures to enhance recovery, while passive restoration allows natural regeneration after pressure removal.

Wetlands' role on carbon pathways towards a net removal of carbon, namely $CO_2$ from the atmosphere and soil/sediment organic matter, encompass three major pathways: i) carbon pool in plant living biomass (% or g C $g^{-1}$ dry weight) closely linked to photosynthesis, as energy source for biomass building blocks (in coastal wetlands this represents a dynamic and smaller amount but ecologically important); ii) carbon sequestration (g C $m^{-2}$ $yr^{-1}$ or t $CO_2$e $ha^{-1}$ $yr^{-1}$) refer to the annual net balance of added carbon per unit area; and iii) carbon stock (Mg C $ha^{-1}$), which correspond to the long-term storage of carbon mostly in sediments (plants rhizosphere) but also carbon accumulated in biomass (above- and belowground) per unit area. These pathways represent meaningful means to offset global carbon emissions ($CO_2$, $CH_4$) [7,8].

To effectively evaluate restoration potential in recovering ecosystem functions and services, accurate, high-resolution mapping of European coastal habitats, ecosystem services, human activities, and pressures is essential. Such mapping underpins evidence-based conservation and restoration planning aligned with global and EU policy targets. At the EU level, detailed habitat maps are critical for meeting reporting requirements under the EU Habitats Directive [9], the EU Water Framework Directive (WFD) [10], and the Marine Strategy Framework Directive (MSFD) [11]. These maps ensure standardized habitat classification and cross-border comparability across Member States, supporting the integrated marine and terrestrial management across EU coastal zones.

Consistent mapping also supports monitoring habitat transitions, adaptive management, and the integration of local restoration into EU-wide strategies. Maps also function as communication tools for assessing policy options, identifying potential co-benefits, and recognizing trade-offs [12]. As European coastal wetlands increasingly face pressures from global environmental change [13], robust habitat mapping provides a scientific foundation for implementing policy targets. Because coastal wetlands are complex socio-ecological systems, mapping should not solely focus on biophysical aspects (supply side of ecosystem services). It must also include the stakeholders, land uses, and activities that shape the demand for ecosystem benefits. This integrated, spatially explicit approach allows for more comprehensive policy-relevant analyses [14]. Spatially explicit datasets encompass biophysical and ecological information from remote sensing and field surveys, ecological status indicators (consistent with EU Habitats and WFD Directives), carbon stocks, and greenhouse gas (GHG) fluxes. Thus, mapping European coastal wetlands must integrate three dimensions: (1) habitats, (2) ecosystem services, and (3) activities and pressures. This triad is a scientific necessity and a policy requirement. For each wetland type, three conditions were initially identified: well-preserved, impacted, and restored. The datasets also include geospatial information from satellite imagery on land cover, vegetation condition, and wetland extent. Remote layers further support mapping of land use, activities, and pressures. These layers are complemented by social datasets capturing perceptions, attitudes, and preferences of stakeholders, landowners, and local communities toward wetland restoration. Such spatially explicit social datasets may be enriched by qualitative information on social acceptability, barriers, motivations, and co-creation processes developed within a Community of Practice (CoP). A CoP is a self-organized group of experts, policymakers, and practitioners working collaboratively to enhance shared knowledge and access to expertise on a specific topic or focus area [16]. A CoP for Wetland Restoration represents a strategic knowledge mobilization framework for scaling coastal wetland restoration, reflecting a deliberate, multi-methodological approach to synthesizing lessons and entrenching them within a functional network [15].

This integrated and harmonized spatial information underpins the NRR ambitions, ensures compliance with multiple directives, and provides the decision-support required to safeguard coastal wetlands of Europe for current and future generations. This information is of high added value when ingested and used in decision support tools that aim at supporting wetland managers and policy makers to prioritize restoration actions [16]. Standardized nomenclatures and classifications allow for the development and application of standardized methods to map and assess habitat extent, potential ecosystem service delivery, and estimate the impact of restoration actions.

In this context, RESTORE4CS, a Horizon-funded project, establishes an integrative socio-ecological framework for European coastal wetlands management and restoration. It involves active participatory methods to identify restoration pathways with stakeholders, uniting field surveys, expert validation, remote sensing, and policy frameworks. It underpins climate mitigation targets and the implementation of EU ecological legislation by delivering decision-support tools and standardizing interoperable, high-resolution habitat maps; quantifying and mapping the supply of multiple ecosystem services; and assessing human activities and pressures to prioritize restoration and inform policy implementation. To this end, six EU case pilots (CPs) were selected considering their broad geographical distribution and ecological representation. Each CP included three comparable conditions (well-preserved, impacted, and restored); and socio-ecological approaches with stakeholders mapping and engagement were already in place. Each selected CP brings added value to scale up at EU level: Ria de Aveiro (Atlantic) – Coastal lagoon, important for biodiversity and carbon storage. Supports understanding of Atlantic wetland restoration. Marjal dels Moros (Mediterranean) – Brackish marshes, offering a model for restoration and management under Mediterranean conditions. Camargue (Mediterranean) – Extensive wetland. Provides insights into restoration in large, complex deltaic systems. Southwest Dutch Delta (Atlantic) – Intertidal area, essential for flood mitigation and biodiversity. Represents restoration in highly managed coastal systems. Curonian Lagoon (Baltic) – Largest European lagoon. Exemplifies restoration challenges and opportunities for Baltic coastal wetlands. Danube Delta (Black Sea) – Europe's largest continuous marshland. Provides a unique perspective on large-scale restoration.

Grounded in RESTORE4Cs main objective, to assess the restoration potential of coastal wetlands in recovering ecosystem functions and associated services related to carbon pathways, and other co-benefits, this paper outlines the different sources of datasets gathered and/or generated within the project, enabling other regions to follow the rational, check data sources, and compare results. Staring with the CPs' description, the sources of the different datasets and how they are combined in a standardized way follows. This paper showcases the added value of these six case pilots supporting the development of restoration strategies and policies based on real-world outcomes.

## 2. Material and Methods

### 2.1 Case Pilots

The dataset covers six coastal wetlands (CPs) across European biogeographical regions (Table 1), representing specific wetland types. Their collective added value supports the integration and upscaling at the EU level.

*Table 1: Description of CPs and respective habitats of interest for restoration*

| Case Pilot | Country | European Basin | River Basin | Coordinates | Interest for Restoration | | | | | | References |
|---|---|---|---|---|---|---|---|---|---|---|---|
| | | | | | **Wetland type** | **EUNIS Level 2** | **Area (ha)** | **Justification for Selection** | **Type of Disturbance** | **Type of Restoration** | |
| Ria de Aveiro | Portugal | Atlantic Ocean | Vouga River | 40.63°N–40.74°N; -8.72°E–-8.59°E | Intertidal seagrass beds (*Zostera noltii*) | A2.7 | ~11,000 | Priority for seagrass recovery; nursery habitat; ongoing restoration actions; priority for local community | Morphological: Erosion of lagoon shoreline, sediment instability, trampling by humans, bioturbation from bait digging; increased turbidity and habitat fragmentation due to eco-hydrological changes. | Active: Transplantation of *Zostera noltii* and physical protection (coconut mats, structures). Passive: Removal of disturbance factors allowing natural sediment stabilization and recolonization. | [17–20] |
| Marjal dels Moros | Spain | Mediterranean Sea | Jucar River | 39.61°N–39.64°N; -0.28°E–-0.24°E | Coastal brackish marshes | A2.5 | ~620 | Represents Mediterranean wetlands under peri-urban and industrial pressures | Hydrological: Drainage, groundwater extraction, artificial flooding. Trophic: Eutrophication from agricultural runoff and wastewater. Morphological: Soil degradation and land-use change. | Mixed: Soil reconstruction, hydrological improvements, active planting of native vegetation, and passive recolonization after pressure reduction. Periodic mowing for habitat heterogeneity. | [21] |
| Camargue | France | Mediterranean Sea | Rhone River | 43.32°N–43.67°N; 4.43°E–4.92°E | Freshwater marshes and ponds | C1 | ~190,000 | Complements saline sites; freshwater restoration under agricultural pressures | Hydrological: Disconnection from natural flooding due to embankments. Land-use: Agricultural intensification (rice fields), irrigation, drainage, fertilizer use causing eutrophication. | Active: Hydrological reconnection, removal of drainage infrastructure, soil and vegetation recovery, topographic reshaping, adaptive management (seasonal flooding, organic farming). | [22] |
| SW Dutch Delta | Netherlands | North Sea | Scheldt River | 51.27°N–51.68°N; 3.99°E–4.40°E | Intertidal salt marshes | A2.5 | Large delta system | Carbon-rich salt marsh; innovative restoration in engineered coastal landscapes | Morphological: Coastal squeeze due to fixed embankments, hard structures (stone breakwaters, wooden pales) altering hydrodynamics, preventing marsh migration, accelerating erosion. | Active: Managed realignment (breaching seawalls to create intertidal habitat), pioneer planting (Spartina anglica with coconut mats), restoring tidal patterns and sedimentation processes. | [23] |

| | | | | | | | | | | |
|---|---|---|---|---|---|---|---|---|---|---|
| Curonian Lagoon | Lithuania | Baltic Sea | Nemunas River | 54.87°N–55.71°N; 20.51°E–21.26°E | Submerged/emergent plant beds | C1 / A2 | ~158,400 | Large-scale lagoon; nutrient reduction and water quality recovery | Trophic: Severe eutrophication from nutrient loading (agriculture, wastewater), cyanobacterial blooms, internal phosphorus release under hypoxic conditions, sediment resuspension reducing light. | Passive: Reduction of external nutrient inputs (improved wastewater treatment, agricultural management), natural recolonization of submerged vegetation, sediment stabilization. | [24–28] |
| Danube Delta | Romania | Black Sea | Danube River | 44.80°N–45.47°N; 28.28°E–29.74°E | Freshwater lakes with reed beds | C1 | ~580,000 | Biodiversity hotspot; large-scale hydrological restoration potential | Hydrological: Disconnection from river flooding due to dikes, drainage for agriculture. Morphological: Soil alteration, reed bed destruction from heavy machinery, habitat loss. | Passive: Hydrological reconnection (dike removal, canal restoration), rewetting former agricultural land, natural recolonization of reed beds and aquatic vegetation. | [29,30] |

**Ria de Aveiro, Portugal**

- **CP description:** Ria de Aveiro is a shallow coastal, covering approximately 11,000 hectares. The lagoon resulted from sea retreat and subsequent coastal strand formation, becoming part of the Vouga River estuary, which hydrographic basin covers approximately 3,362 $km^2$. It represents a complex socio-ecological system within the Natura 2000 network, including freshwater marshes, a natural freshwater lake, riparian vegetation along the main river and tributaries, salt marshes, seagrass meadows, mudflats, sandbanks, dunes, and sand beaches. The Ria de Aveiro Natura 2000 area encompasses three aquatic ecosystem domains: freshwater, transitional waters, and coastal marine waters. The transitional water system corresponds to the lagoon system connecting the Vouga basin (source of 80% of its freshwater) to the Atlantic Ocean through a single inlet, with hydrodynamics forced by tidal action. *Zostera noltei* meadows and one of the largest continuous salt marshes in Europe, with populations of *Sporobolus maritimus* and *Juncus maritimus*, supports the biological richness of the system. The territory is mostly located in the Mediterranean biogeographical region, with the northern end in the Atlantic biogeographical region, characterized by Atlantic Ocean influence with maritime temperate characteristics.
- **Well-preserved condition:** Intertidal seagrass beds with continuous coverage, healthy sediment structure, and absence of significant pressures such as erosion, bioturbation, bait digging, or invasive species presence. Healthy intertidal seagrass population in the vicinity of restored and/or altered conditions were selected. Natural vegetation and substrate conditions indicated minimal disturbance and good ecological status.
- **Altered condition:** Bare, unvegetated intertidal areas where *Z. noltei* meadows have been lost or with severely fragmented populations. These areas exhibit erosion, unstable sediments prone to resuspension and reduced biodiversity. The sediments show signs of bioturbation from bait-digging activities and physical disturbance from trampling. Nutrient fluxes at the sediment-water interface are altered, with increased release of nitrogen (N) and phosphorus (P) compared to vegetated areas [31].
- **Restored condition:** Active vegetation transplantation of *Z. noltei* has been ongoing in areas where pressures are no longer relevant, to mitigate erosion and stabilize sediments. Mosaic-transplants were able to cover previously unvegetated areas and develop uniform, robust coverage within one year. Physical protection measures were applied to seagrass transplants in areas with relevant invasive species. The restoration demonstrated feasibility of seagrass transplantation in degraded and contaminated ecosystems. Plant-mediated biogeochemical processes at transplanted areas were found to reduce nutrient fluxes at the sediment-water interface, with restoration of N and P regulation functions occurring no later than one year after transplantation [31]. Passive restoration also occurred where disturbance factors were eliminated, allowing natural recovery of sediment stability.

**Marjal dels Moros (Valencian Wetlands), Spain**

- **Description:** Located in Puçol and Sagunt (Valencia, Spain), the CP covers about 620 ha owned by the Generalitat Valenciana and private owners, with roughly two-thirds devoted to environmental conservation. In the 1990s the Generalitat acquired the area and launched a wetland restoration project to recover natural conditions after decades of agriculture, drainage, and degradation. The wetland hosts high biodiversity, valuable flora and fauna, and notable hydrological and geomorphological features. It lies in a peri-urban setting, surrounded by buildings and crops, and subject to strong agricultural and industrial pressures. Water inputs include rainfall, wastewater, agricultural drainage, controlled well pumping, and seawater intrusions via groundwater rather than open marine exchange. Its Holocene evolution reflects Plio-Quaternary deposits from the Sierra de Calderona, sedimentation from the Palancia River, and the formation of a coastal-barrier lagoon system. Historical research documents Roman and Muslim occupation (1st–13th centuries) and their irrigation networks. In the High Middle Ages, population growth and agricultural expansion—especially rice cultivation—triggered initial hydrological changes. In recent decades, the wetland has faced pollution from industry and wastewater, aquifer overexploitation, and invasive species such as *Arundo donax*. Until 2000 the area was used for farming and hunting, later prohibited after its acquisition and the implementation of restoration and protection measures.
- **Well-preserved condition:** Coastal brackish marshes with intact emergent swamp communities, stable hydrological connectivity (marine intrusion via groundwater), and limited structural alteration or water quality degradation. These areas featured functional wetland habitats with sustained native biodiversity, appropriate water levels maintained through natural rainfall, groundwater sources, and controlled seawater intrusion. The vegetation consisted of plant communities adapted to brackish conditions with variable salinity regimes, including reedbeds, bulrush stands, and halophytic shrub communities.
- **Altered condition:** Degraded marshes altered due to hydrological changes (drainage, groundwater extraction, and artificial water supply from irrigation and wastewater sources leading to partial desalinization), trophic alterations (nutrient enrichment from agricultural runoff, wastewater, industrial pollution), and morphological modifications (land use changes, soil degradation). These areas experienced habitat loss, reduced native vegetation, proliferation of invasive species, degraded water quality with elevated nutrient concentrations, and loss of characteristic brackish marsh ecological functions because the disruption of the natural salinity gradient. Agricultural exploitation and hunting activities had significantly impacted vegetation structure and ecosystem services until 2000.
- **Restored condition:** Active restoration included soil reconstruction to improve substrate conditions, morphological modifications including topographic adjustments to restore appropriate elevation and hydrological connectivity, and active planting of native vegetation. Hydrological improvements involved ensuring good condition and functioning of hydraulic infrastructure, diversifying water supply sources including reuse of irrigation surpluses and controlled groundwater resources and maintaining appropriate flood levels even in dry years to ensure aquatic refuges. Vegetation recovery involved both active planting of native species and passive recolonization after reducing external pressures and stabilizing water regimes. Mowing of helophytic vegetation with amphibious machines is regularly implemented as a management measure.

**Camargue, France**

- **Description:** Located along the Mediterranean coast in southern France, the Camargue forms the principal territorial unit of the Rhone delta, covering over 190,000 hectares. This extensive floodplain supports a diverse mosaic of habitats, including lagoons, freshwater and brackish marshes with emergent or submerged vegetation, halophytic scrubs, and steppe-like grasslands. These natural environments are interspersed with agrosystems dominated by irrigated crops. A complex network of irrigation and drainage channels underpins the hydrological functioning of the delta. Approximately 730 million $m^3$ $y^{-1}$ of water are pumped from the Rhone to compensate for river embankment, prevent soil salinization, and sustain primary productivity during the summer drought period. Around half of this volume returns to the Rhone through drainage canals, while the remainder flows towards the main lagoon system, the Vaccarès. Although primarily managed to support rice cultivation, this water is also used for the flooding of marshes, reed harvesting, wildfowl hunting, and irrigation of wet meadows. Rice cultivation has historically played a pivotal role in shaping both the economy and ecology of the Camargue, contributing to soil desalination, freshwater maintenance for agricultural practices, and providing critical habitat for migratory birds. The Camargue represents a tightly interlinked socio-ecological system, where water management practices are central to balancing agricultural productivity, biodiversity conservation, and wetland resilience.
- **Well-preserved condition**: Freshwater marshes and ponds in the Camargue that retained natural hydrological regimes and ecological features, without significant historical land use conversion or hydrological alterations. These areas maintained intact soil, vegetation, and water flow supporting natural wetland functions. Natural habitats featured emergent or aquatic vegetation with seasonal flooding patterns, supporting native flora and fauna adapted to brackish seasonal conditions.
- **Altered condition:** Former rice fields and fishponds. These areas experienced fundamental hydrological modifications through physical modifications, artificial irrigation and drainage systems, soil alterations from agricultural practices, pesticide application affecting biodiversity, and loss of native vegetation. The natural seasonal hydrological variability was replaced by managed water regimes, with continuous flooding during dry seasons.
- **Restored condition:** Former rice fields and pastures that underwent soil, hydrology, vegetation, and morphological reconstruction are considered as restored habitats. The Cassaïre estate restoration project converted around 70 hectares of former agricultural land and hunting ponds into a mosaic of natural meadows and temporary marshes representative of Mediterranean wetlands. Implemented by the Tour du Valat and the Marais du Vigueirat, the project aimed to restore ecological functions through topographic reshaping, removal of drainage and irrigation infrastructure, and the re-establishment of seasonal flooding regimes typical of the Camargue. Soil and seed transfers were used to accelerate natural recolonization by native species, while extensive grazing was maintained to promote habitat heterogeneity. Monitoring has shown the rapid recovery of wetland vegetation and the return of rare and indicator species. The Petit Badon restoration project, launched in 2020 on former rice fields, involved re-shaping the terrain and installing water control structures to re-establish natural flooding and drying cycles. Adaptive management measures such as organic farming, dry sowing, and seasonal flooding were introduced to promote biodiversity and restore ecological functions. Early monitoring has shown rapid recolonization by wetland vegetation and increasing use of the site by amphibians and waterbirds.

**Southwest Dutch Delta (Oosterschelde/Westerschelde Wetlands), Netherlands**

- **Description:** The South-West Delta contains natural values rare in Europe, with high species diversity and habitat for large numbers of species. A severe north-westerly storm in 1953 combined with spring tides flooded large parts of the South-West Delta in the Watersnoodramp, the largest Dutch natural disaster of the 20th century. After the disaster, the Delta Works restored water safety through dams and storm surge barriers that greatly shortened the coastline and limited the dynamics of sea and rivers and fresh and salt water, creating different water basins (fresh and salty, with and without tides). In 1986, the Eastern Scheldt (Oosterschelde) was closed off from the sea by a storm surge barrier which still allows tidal action. As a result of tidal currents, erosion and sedimentation processes occur, resulting in variable patterns of salt marshes, mudflats, tidal flats (the intertidal zone), shallow water, and deep tidal channels. Large areas of mudflats occur in the east and north of the area. Inside the dike, the area between outer and inner dikes (*inlaag*) provides additional protection against flooding and consists mainly of wet grasslands and open water. The marshland plays a major role in flood mitigation. The Western Scheldt (Westerschelde) is a waterway in open communication with the North Sea and the Scheldt, serving as an important link between Antwerp, Belgium, and the North Sea, one of the world's busiest waterways with large tidal ranges. The Eastern Scheldt has no longer a direct connection to the river Scheldt, and therefore no freshwater inputs from the river.
- **Well-preserved condition:** Intertidal salt marshes showing stable hydrological and sedimentation processes, vegetated surfaces, and minimal disruption by coastal infrastructure (e.g., absence of stone breakwaters or wooden pales). These areas maintained natural marsh integrity and ecological processes including tidal inundation patterns, natural sedimentation, pioneer zone species, and mid-upper marsh communities. The wide variety of environmental types (determined by tides, currents, water temperature, altitude, water quality, and sediment composition) was accompanied by great diversity of animal and plant species.

- **Altered condition:** Areas where stone breakwaters or wooden pales perpendicular to the marsh were installed to reduce hydrodynamics and locally reduce erosion. These hard structures disrupted natural sedimentation processes, reflected wave energy, and altered natural marsh development patterns. Coastal infrastructure caused accelerated erosion through modification of hydrodynamics and prevention of landward marsh migration. Pioneer-zone species disappeared as a result of lateral erosion. Areas experienced loss of vegetated surfaces and increased erosion rates due to fixed coastal defenses.
- **Restored condition:** Restoration focused on morphological reconstruction and hydrological recovery through managed realignment. As restored habitats, areas were selected where: (a) managed realignment was done to create marsh habitat in 2019; and (b) unintended managed realignment resulting from a dike failure in 1990 that ended up creating marsh habitat. Managed realignment involved moving the coastal defense line further inland by breaching existing seawalls to provide new intertidal habitat inland, allowing salt marsh to develop on previously reclaimed land. Pioneer salt marsh restoration using *Spartina anglica* plants fixed in coconut mats was applied at locations in the Eastern Scheldt to investigate the potential of this method to re-establish pioneer saltmarsh. Under the right conditions (e.g., sufficient sediment input), the *Spartina* plants grow out and form larger tussocks and subsequently meadows, adding to coastal protection, biodiversity, and ecosystem functioning. Vegetative cover rapidly develops following sedimentation at managed realignment sites. Restoration facilitated recovery of natural hydrodynamics, tidal patterns, sedimentation processes, and vegetation establishment.

**Curonian Lagoon, Lithuania**

- **Description:** The Curonian Lagoon is the largest coastal lagoon in Europe with a total area of 1,584 $km^2$. It is a shallow waterbody with an average depth of 3.8 meters. The lagoon is transitional water body between the south-eastern Baltic Sea and the watershed of the Nemunas River, one of the largest river systems in the south-eastern Baltic Sea and contributes most of the water, sediments, and nutrients loads to the Curonian Lagoon. The Nemunas is one of the four most important rivers in the Baltic Sea region and the main freshwater input, discharging nutrient-rich waters in the central part of the lagoon. The river inflow affects water circulation patterns [32] and results in different water renewal times between the northern and the southern part of the lagoon. Thus, the northern part is a transitional, estuarine system predominantly flushed by freshwater with occasional brackish water inflows. The central-southern part is functioning as a lacustrine-like system, being highly eutrophic with frequent cyanobacteria blooms in late summer, when chlorophyll a concentration can reach up to 400 mg $m^{-3}$ [33]. The lagoon hydrodynamics shape the sedimentary environment, consisting of a mosaic of sand, silt, mud, and shell deposits [34–36]. The shores and shallow littoral of the lagoon are colonized by emerged (*Phragmites australis)* and submerged macrophytes (e.g., *Chara* spp., *Myriophyllum* sp., *Stuckenia* sp., *Potamogeton perfoliatus*). In these areas, two dominant species (*Chara contraria* and *Chara aspera*) form monospecific meadows [28]. The Nemunas Delta is a managed nature reserve, regional park, Natura 2000 site, and is also recognized as RAMSAR wetland. The delta comprises marshes, raised bogs, ridge-pool complexes, flooded forests, and meadows, supporting a rich plant community with numerous endangered species. Located on the East Atlantic flyway, it is an internationally important breeding, wintering, and stopover site for thousands of waterbirds and migratory birds.

- **Well-preserved condition:** characterized by high coverage of submerged aquatic vegetation, sandy or mixed bottom substrates. These areas exhibited strong ecological integrity and lower phytoplankton productivity, with mean chlorophyll-a concentration not exceeding 40 mg $m^{-3}$ during the summer months [33]. The submerged vegetation provides habitat, stabilizes sediments, and contributes to nutrient cycling under more balanced trophic conditions [28]. Light availability in these habitats is sufficient to support submerged macrophyte growth, which reinforces water clarity and sediment stability.
- **Altered condition:** characterised by accumulated muddy sediments and reduced coverage of submerged aquatic vegetation. In these altered areas, mean summer chlorophyll-a concentrations typically exceed 60 mg $m^{-3}$, and cyanobacterial surface scums, dominated by *Aphanizomenon flos-aquae* and *Microcystis* sp. occur frequently [33]. Elevated autotrophic and heterotrophic respiration can deplete bottom water $O_2$, resulting in short-term hypoxic or anoxic events and release of reactive P from sediments [26,27]. During such periods, phytoplankton primary production exceeds mineralization capacity of the system, by 60%, leading to the accumulation of organic matter in water column and surface sediments. In these areas, high turbidity and sediment resuspension limit light penetration to the sediments, preventing recolonisation by submerged vegetation and affecting cyanobacterial dominance.
- **Restored condition:** Restoration actions focused on improving water quality. Improvements of wastewater treatment infrastructure were implemented reduced point source pollution, moreover nutrient load reduction in the Nemunas River watershed was achieved through decreased fertilizer use and industrial production controls. Restored areas are characterised by habitats by reduced mud accumulation and recently increased submerged vegetation cover and maximum depth of colonisation [28]. The recovery of sandy substrate areas and expansion of submerged aquatic vegetation likely resulted from improved water quality conditions in combination with reduced turbidity and enhanced high availability. Improved water quality reduced phytoplankton biomass and suspended solids, increasing light penetration to the sediments and enabling recolonisation by submerged vegetation.

**Danube Delta, Romania**

- **Description:** The Danube Delta encompasses over 5,800 $km^2$ and represents one of the best-preserved deltas globally, hosting a mosaic of ecosystems supporting significant biodiversity. The Delta includes over 300 lakes ranging in size between 14 and 4,530 hectares, with water depths of 1.5 - 4 m. The extensive reed beds constitute the largest compact reed bed expanse worldwide, playing significant roles in water purification, soil formation and stabilization, biomass production, and carbon storage. Seasonal flooding of the Danube, with higher water levels in spring and low water levels in autumn, shapes the delta's terrain and influences species distribution and abundance. The hydrological regime creates temporary aquatic habitats rich in nutrients that are vital for numerous species' life cycles.
- **Well-preserved condition:** Freshwater lakes and reed bed complexes that preserved natural hydrology (permanent or seasonally inundated), native vegetation cover, and undisturbed morphology. This area, located on the right bank of the Danube between Isaccea and Tulcea, covers 9,170 hectares and comprises a network of lakes and swamps interconnected by small canals, representing a remnant of the Danube floodplain under semi-natural flooding regime. The aquatic habitats consist of freshwater shallow lakes with submerged vegetation (*Potamogeton spp.*, *Ceratophyllum spp.*) or floating vegetation (*Trapa natans*, *Nymphaea alba*) surrounded by reed beds, supporting high biodiversity and important fishery resources. Selected preserved sites include Lake Gorgonel (141 hectares) and Tilincea (188 hectares), both with maximum depths of 3 meters and classified as good ecological status according to the WFD.
- **Altered condition:** Former freshwater wetlands converted to dryland during the 1980s. One site is located near Tulcea city in the former Danube floodplain and is currently used for agriculture, mainly cereal production. The second area, near Mahmudia city and called Carasuhat, was initially intended for agricultural use but proved unsuitable for crops and was subsequently used as pasture for cattle. These areas lost their main structural and functional ecological attributes through disruption of the hydrological regime, elimination of native vegetation, and soil alteration. During project implementation, Carasuhat altered area was accidentally flooded due to the breakage of the dike isolating the area from the river network, and the area has been abandoned (no use for cattle pasture) since.
- **Restored condition:** The first restored site, Zaghen Lake near Tulcea city, underwent restoration aimed at increasing open water area by reducing reed surface while maintaining a hydrological regime connected to the Danube River. The hydrological regime is maintained through a canal connected to the river and, in cases of extremely low water levels, with the assistance of a pumping station. Before restoration, the area was disconnected from the Danube's hydrological regime. Restoration activities finished in 2015, with a total restored surface of 250 hectares. The second area, Carasuhat, consists of a former pasture restored to wetland by restoring the hydrological regime. A total area of 924 hectares of farmland was converted into a wetland ecosystem starting in 2016, transforming the area into a rich habitat for numerous bird species and a popular spot for birdwatching.

## 2.2 Biophysical and ecological data

Habitat mapping was conducted using the hierarchical EUNIS habitat classification database. Initial habitat data were extracted from the EUNIS shapefiles available on the European Environment Agency's (EEA) EUNIS portal, which provides comprehensive spatial data for standardized European habitat types [37]. Subsequently, local experts refined the preliminary classifications by integrating regional databases and historical surveys to enhance accuracy and ecological relevance. The data was then harmonized and updated to EUNIS terrestrial 2021 [38] and marine 2022 [39], using the tabular datasets with crosswalks, and the local experts did a final validation, eliminating ambiguity caused by multiple matches. This multi-step approach assured methodological consistency and allowed integration of updated hierarchical habitat categories while maintaining coherence with prior nomenclatures and facilitating interoperability with European environmental reporting obligations.

## 2.3 Socio-ecological data

For each EUNIS landscape unit, ecosystem services were systematically assigned based on the Common International Classification of Ecosystem Services (CICES) version 5.1 framework [40]. The CICES nomenclature encompassed provisioning services, regulation and maintenance services, and cultural services, and the local experts were responsible to assign the ecosystem services provided by each habitat in their CP, ensuring ecological accuracy and contextual relevance.

The methodology proposed by Borgwardt et al. (2019) to categorize activities and pressures was followed. Anthropogenic activities and pressures were identified through satellite imagery for land cover, vegetation health, and wetland extent complemented with remote sensing layers and systematic field surveys, literature review, and expert consultation, aligned with Habitats Directive, WFD, MSFD, and the statistical classification of economic activities (NACE - *Nomenclature Statistique des Activités Économiques*) codes. The dataset captures a detailed record of direct and indirect effects of natural and anthropogenic activities and pressures on the wetland functionality.

## 2.4 Stakeholder and community data

A multi-layered participatory approach was followed to integrate stakeholder and community perspectives into coastal wetland restoration assessment across CPs. This participatory framework was designed to capture local stakeholder preferences, perceptions, and priorities regarding restoration actions, ecosystem services valuation, and decision-making processes - particularly focusing on carbon pathways and associated co-benefits. The participatory methodology relied on a structured multi-criteria analysis (MCA) framework that positioned stakeholders as active co-designers rather than passive information sources.

The MCA framework proceeded through three integrated steps: Step 1) contextual scoping through semi-structured interviews with CP leaders to characterize site boundaries, ecosystems, activities, pressures, and stakeholder networks, while simultaneously developing preliminary criteria lists. Step 2) engage workshop participants in structured criteria weighting exercises on a 1-9 importance scale, with 1-2 hours dedicated to weighting embedded within 3–4-hour sessions that included contextual briefings and moderated discussions, explicitly distinguishing between criterion importance in decision-making and projected impact magnitude. Steps 3) operationalized each criterion through quantitative, qualitative, or monetary indicators sourced from in-field measurements (GHG, habitat extent, hydrological assessments), administrative records (employment, tourism, agricultural data), stakeholder knowledge through expert elicitation, and custom survey instruments, subsequently enabling indicator evaluation under alternative management scenarios.

The knowledge base supporting the ECoP was built through multiple parallel data streams. The first phase involved a scoping review of scientific literature, examining fifty peer-reviewed publications from repositories including ResearchGate, Elsevier, and JSTOR. These sources were prioritized based on geographical scope (European focus), publication currency, and institutional links to European research programs (Horizon, LIFE, Interreg). This literature review identified both barriers and enablers of large-scale restoration, providing theoretical grounding for subsequent case study selection and comparative analysis. Thirty-seven interviews conducted with stakeholders from these pilot sites, providing insights into both historical restoration initiatives and contemporary adaptive management approaches within each coastal wetland system. Complementing these interviews, six participatory workshops (the same held for the MCA, to avoid stakeholder fatigue) were organized during the second semester of 2024 and early 2025.

The engagement at regional scale represents a critical strategic mechanism for extending methodologies and findings beyond the six pilot sites to broader geographical and administrative scales. To extend the community beyond CP boundaries, interviews were conducted with twenty partners representing civil society organizations, non-governmental organizations, international organizations, government agencies, private foundations, and research institutions engaged in large-scale wetland restoration across Europe. Simultaneously, the ECoP engaged with fourteen European projects working on wetland restoration, creating a distributed collaborative network distinct from formal institutional structures. The knowledge base supporting the ECoP recruitment was further strengthened through systematic literature review and grey literature analysis. This included fifty scientific publications sourced from academic repositories, and thirty-four additional publications encompassing national reports from local, subnational, and national authorities, Interreg project websites, Ramsar Convention materials, wetland-based solutions factsheets, and LIFE programme project entries. These multiple recruitment pathways - direct stakeholder engagement at pilot sites, extended network interviews, inter-project collaboration, and comprehensive literature synthesis- created a heterogeneous community representing diverse organizational types, biogeographical regions, wetland types, and restoration approaches across the European space.

## 3. Results and Discussion

Decision support and prioritization tools are crucial for national/regional restoration plans. Effective, scalable tools require standardized, harmonized datasets for baseline mapping, pressure/threat analysis, and ecosystem service effects estimates. The described datasets largely meet these needs.

Applying the EUNIS habitat classification across CPs provides the standardization mechanism essential for effective continental-scale restoration planning. Standardized habitat mapping is vital as past assessments were hampered by heterogeneous classification systems, limiting synthesis of outcomes. Harmonization to EUNIS terrestrial 2021 and marine 2022 ensures compatibility with European environmental reporting obligations, facilitating integration with existing and coming regulatory frameworks. EUNIS's hierarchical structure, from broad habitat groups to fine-scale ecological units, offers analytical flexibility for multi-scale questions, from site-specific work to continental policy [41]. This multi-scale analytical capacity responds to policy demands emerging from the NRR, which requires Member States to submit detailed national restoration plans with prioritization strategies for diverse coastal wetland ecosystems [6].

A total of 38 distinct habitats at EUNIS level 2 were identified in the six CPs (Figure 1), covering diverse natural and human-shaped coastal wetland ecosystems such as intertidal salt marshes, seagrass meadows, coastal lagoons, freshwater marshes, inland water bodies, saltpans, and rice fields. Notable habitats include extensive littoral sediments in the Camargue, Southwest Dutch Delta and Curonian Lagoon sites (relative cover up to 85-95%), intertidal seagrass beds and coastal dunes in Ria de Aveiro, freshwater marshes and reedbeds in the Danube Delta, and arable land in Valencian Wetlands [42]. The legal protection status and policy instruments such as Natura 2000 and EU Habitats Directive Annex I apply variably across habitats and pilots.

Recent advances demonstrate that ensemble machine learning models combined with high-resolution satellite imagery and ecologically meaningful environmental variables can produce European habitat maps with independent validation and uncertainty analyses, significantly improving both thematic and spatial resolution compared to traditional manual classification approaches. Integrating the hierarchical nature of EUNIS classifications through deep learning frameworks substantially improves classification accuracy for Level 3 habitat types across heterogeneous European ecosystems. This technological integration with standardized habitat frameworks provides an operational pathway to scale point-scale habitat data from CPs to continental assessments, ensuring that fine-scale ecological knowledge can inform landscape-level policy implementation [43,44].

For the socio-ecological data, the successful linkage between EUNIS habitat types and CICES v5.1 ecosystem service categories operationalizes the ecosystem service cascade model, enabling quantitative estimation of service supply from spatial habitat maps. The direct crosswalk between habitat classification and ecosystem service frameworks is a key advancement for restoration prioritization, allowing managers to predict how habitat-specific restoration actions translate into changes in provisioning, regulating, and cultural services. Previous studies demonstrate that ecosystem services vary substantially among coastal wetland habitat types, emphasizing the importance of habitat-level service assessments over aggregated valuations [45,46].

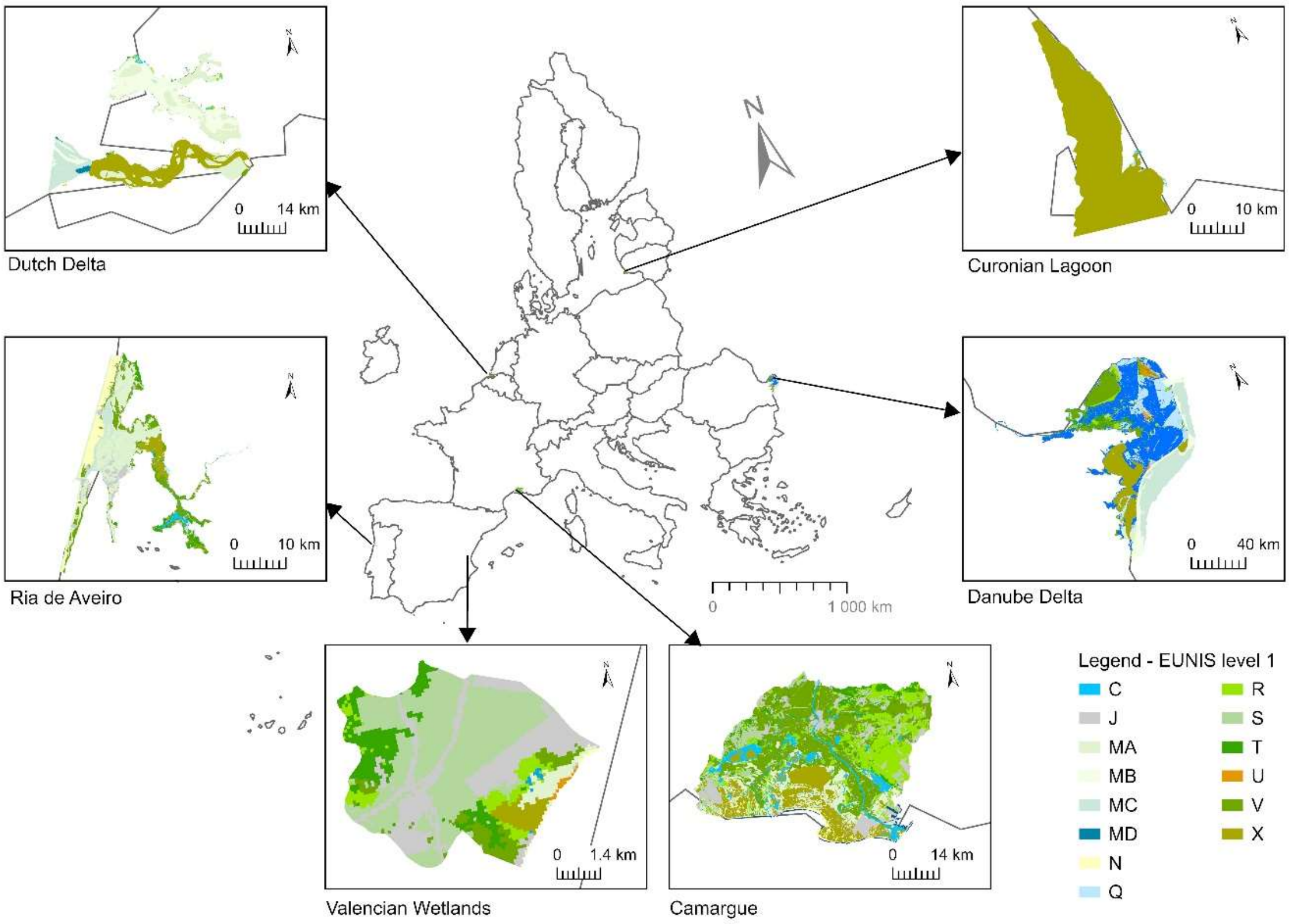

*Figure 1: Case Pilot location and EUNIS level 1 habitats. Spatial data were projected in WGS 1984 geographic coordinate system (EPSG:4326). Country administrative areas from GISCO, European Commission.*

The dataset encompasses 1,612 records linking habitats to ecosystem service classes under the CICES v5.1, providing comprehensive coverage of provisioning services (including biomass production for nutrition, materials, and energy; genetic materials; and non-aqueous abiotic outputs such as wind and solar energy), regulation and maintenance services (water quality regulation, erosion control, atmospheric composition regulation, and lifecycle maintenance including habitat and gene pool protection), and cultural services (intellectual and representational interactions including scientific investigation and education; physical and experiential interactions including recreation and health; and non-use values linked to biodiversity and landscape appreciation) [42]. This spatially explicit ecosystem service cascade recognizes that service provision (supply side) must be evaluated in relation to human needs and benefits requests (demand side), addressing a critical analytical gap in restoration planning. Studies show supply and demand are frequently spatially mismatched, with supply concentrated in rural and less-developed areas while demand is concentrated in urban and economically developed zones [47]. By mapping both supply (habitat mapping) and demand (stakeholder engagement), the RESTORE4Cs approach identifies priority zones where restoration can simultaneously address ecological degradation and unmet service demands. This integration supports identifying co-benefits and trade-offs, enabling transparent multi-objective decision-making (demand side) [47].

The ecosystem services flow is mediated by human activities establishing activity-pressure associations that impact the ecological status of habitats and compromise service provisioning. The extensive compilation of 23,064 habitat - ecosystem services – activity - pressure associations detail the anthropogenic activities taking place in the pilots. These are categorized using harmonized groupings derived from European legislative frameworks including Habitats Directive, WFD, and MSFD, supplemented by NACE codes [14]. The dataset distinguishes direct and indirect pressures such as habitat modification, pollution, eutrophication, hydrological alterations, invasive species, and infrastructure development. Explicitly associating these pressures with EUNIS habitats and CICES groups enables targeted impact and restoration priority assessments[14]. This significantly advances conventional planning, which often lacks explicit documentation of causal pathways linking activities to degradation and outcomes.

Regarding stakeholders and community data, across the CPs, workshops assembled between 7 and 29 participants, with individual surveys collecting 7 to 21 valid answers per site. Participants included governmental environmental agencies, municipalities, NGOs, research institutions, traditional resource users, cultural heritage practitioners, tourism operators, and civil society. This diverse engagement reflects the critical principle that coastal wetland restoration decisions involve trade-offs among competing objectives, requiring comprehensive stakeholder input to reach socially acceptable strategies. A key finding was the need for significant linguistic and conceptual adaptation across contexts, stressing that participatory methodologies must be iteratively adapted to local discourses and frameworks for genuine co-production of knowledge. This adaptive approach aligns with contemporary best practices in participatory research, which emphasize that meaningful stakeholder engagement requires not only structured elicitation mechanisms but also recognition and accommodation of how different communities understand and conceptualize environmental problems [48,49]. Participatory multi-criteria analysis literature confirms that stakeholders are rarely involved throughout entire decision-making processes, yet studies on nature-based solutions show that when stakeholders participate across problem definition, criteria development, and prioritization stages, implementation success rates increase substantially [50]. The hierarchical integration of scoping, participatory preference elicitation, and scenario-based assessment creates a transparent, replicable methodology for systematic evaluation of restoration actions' multidimensional impacts across geographically diverse coastal wetland contexts. Coastal wetland restoration decisions inherently involve trade-offs among competing environmental, social, and economic objectives, and decision-makers require comprehensive stakeholder input reflecting diverse values and knowledge systems to reach acceptable and robust restoration [18].

Managers involved in national roadmaps form Communities of Practice (CoP) at regional and EU levels (ECoP), engaging in strategic planning and effective implementation of the Regulation. Initiatives with the ECoP included a workshop at each pilot, an online European workshop, and an autumn school for sixteen policymakers. The ECoP is explicitly designed to capitalize on project learnings and promote their dissemination and replication at regional scales, including scaling application from pilot sites to catchments, national zones, and pan-European levels. This scalability is achieved by standardizing monitoring and prioritization tools developed at pilot sites. The ECoP structure exemplifies how learning networks translate local innovations into regional/EU policy. This approach addresses a gap in restoration science by embedding participatory processes within formal institutional frameworks, creating scaffolding for deeper learning processes (double-loop/triple-loop learning) to emerge and influence policy. As end-users, their feedback was paramount for the usability and pertinence of the technological interface. The participatory methodology demonstrated that rigorous integration requires: (1) structured yet adaptive frameworks accommodating local contexts; (2) multi-layered data integration (ecological, economic, preferences); (3) transparent documentation of indicators/methods; and (4) recognition of subjectivity/power dynamics. These principles ensure replicability while maintaining contextual relevance, crucial for upscaling.

Supported by the integrated datasets, impact and restoration priority assessments can be conducted in a more integrative and holistic manner using spatial decision support systems, which assist in solving spatial problems (e.g., land use decisions) by consider co-benefits and trade-offs between ecosystem services and activity-pressure associations. These systems operationalize spatially distributed data into decision environments by integrating biophysical parameters (habitat condition, degradation pressures, restoration potential), socio-ecological parameters (ecosystem service supply and demand, stakeholder preferences), and governance considerations (regulatory requirements, implementation feasibility, equity concerns). Beyond data integration, the RESTORE4Cs Spatial Prioritization Toolbox operationalize these datasets into a decision-support environment that helps identify and rank potential restoration areas across European coastal wetlands, highly demanded by national and subnational entities to prepare the restoration plans in the context of the NRR. The Toolbox combines harmonized spatial layers on habitats, degradation pressures, and potential ecosystem service supply to compute multi-criteria indices of restorability and restoration relevance, enabling users to visualize where restoration actions can maximize co-benefits for climate mitigation, biodiversity conservation, and water cycle regulation. At the pilot level, the Toolbox ingests fine-scale spatial data to identify priority zones where interventions are most likely to restore functions or reconnect degraded areas. Harmonization and standardization of spatial data ensures seamless visualization and comparison in the RESTORE4Cs data platform.

By documenting the standardized data structures, indicator definitions, and methodological choices used in CPs, the RESTORE4Cs approach adaptable to coastal regions lacking detailed baseline data. Machine learning techniques, including transfer learning, offer pathways to extrapolate pilot-level insights to broader European coastal zones. Additionally, remote sensing technology enables systematic temporal monitoring of restoration effectiveness over large areas, supporting adaptive management. The dataset and tools fulfill a critical need: effective restoration prioritization requires not only rigorous ecological assessment but also explicit integration of ecosystem service valuation, stakeholder preferences, regulatory requirements, and transparent documentation.

## 4. Conclusions

The comprehensive dataset developed through the RESTORE4Cs project, offers a robust, interoperable foundation for advancing European coastal wetland habitat conservation and restoration. These data sets include biophysical and ecological data, socio-ecological data, and data and information on stakeholders and communities' engagement, that is spatial explicit at the level of habitats, as land scape unit. By systematically mapping habitat distributions using updated EUNIS classifications alongside detailed ecosystem service assessments through the CICES framework, the dataset enables a nuanced understanding of the diverse biogeographical and ecological contexts defining these wetlands across Europe. Furthermore, the integration of anthropogenic activities and associated pressures following EU wetlands-related directives standards, provides critical insights into the spatially explicit drivers impacting wetlands ecosystems functions and provided services. This triad of habitat, ecosystem services, and activities and pressures mapping is also essential for stakeholders and citizens engagement, enabling evidence-based decision-making, following a participatory process. This transdisciplinary approach, foster social acceptance of restoration options, allowing for targeted restoration interventions, cross-regional comparisons, and compliance with evolving EU environmental legislation such as the Nature Restoration Law and Biodiversity Strategy 2030.

Beyond its immediate application in site-specific restoration planning, these datasets serve the RESTORE4Cs toolbox to address broader environmental challenges including climate change mitigation, biodiversity loss, and sustainable land management. Ultimately, RESTORE4Cs contributes significantly to advancing science-policy interfaces by sharing spatial explicit data, following FAIR principles, as well as two technological interfaces, Toolbox and Decision Support Tool, that underpin sustainable coastal wetland management and restoration to safeguard these ecosystems for future generations.

## Data Availability

The datasets are publicly available via LifeWatch Italy Data Portal: https://data.lifewatchitaly.eu/collections/7bf0ca24-4992-4a0c-96f3-ef2c4596e9c1?configuration=portals, with CC BY-SA 4.0 license:

- Habitats and Ecosystem Services: DOI https://doi.org/10.48372/F9EQ-5359

- Habitats, Activities and Pressures: DOI https://doi.org/10.48372/ADX2-VJ72

The datasets are CSV-formatted, UTF-8 encoded, compatible with R, Python, ArcGIS, QGIS, and standard databases.

**Author contributions**

**Bruna R.F. Oliveira:** Conceptualization; Data curation; Formal analysis; Writing – original draft; Writing – review and editing

**Ana I. Lillebø:** Conceptualization; Methodology; Investigation; Formal analysis Writing – original draft; Writing – review and editing; Funding acquisition; Project administration

**Antonio Camacho, Anis Guelmami, Christoph Schroder**: Methodology; Investigation; Formal analysis; Writing – original draft; Writing – review and editing; Funding acquisition

**Nina Bègue, Martynas Bučas, Constantin Cazacu, Elisa Ciravegna, João Pedro Coelho, Relu Giuca Constantin, Samuel Hilaire, Marija Kataržytė, Daniel Morant, Antonio Picazo, Nico Polman, Marinka van Puijenbroek, Justine Raoult, Carlos Rochera, Michaël Ronse, Ana I. Sousa; Tudor Racoviceanu, Diana Vaičiūtė:** Investigation; Formal analysis; Writing – review and editing

**Funding**

This research has received funding from the project RESTORE4Cs - *Modelling RESTORation of wEtlands for Carbon pathways, Climate Change mitigation and adaptation, ecosystem services, and biodiversity, Co-benefits* (DOI: 10.3030/101056782), co-funded by the European Union under the Horizon Europe research and innovation programme (Grant Agreement ID: 101056782). The views and opinions expressed are those of the author(s) only and do not necessarily reflect those of the European Union or the granting authority. Neither the European Union nor the granting authority can be held responsible for them.

**Acknowledgments**

We acknowledge the contributions of local experts at each CP and the broader RESTORE4Cs CoP for their expertise in habitat classification and ecosystem service assessment. We acknowledge the contributions of Eleftheria Kampa (policy), Santiago Suarez (community of practice), Clementine Anglada (environmental economics) and Auriane Bodivit (environmental economics). We acknowledge the support of FCT – Fundação para a Ciência e a Tecnologia I.P., to CESAM-Centro de Estudos do Ambiente e do Mar, under the project references UID/50017/2025 (doi.org/10.54499/UID/50017/2025) and LA/P/0094/2020 (doi.org/10.54499/LA/P/0094/2020)

## References

[1] M. Otero, A. Camacho, D. Abdul Malak, E. Kampa, A. Scheid, E. Elkina, How can coastal wetlands help achieve EU climate goals? Policy Brief., 2024.

[2] IPBES, The global assessment report of the intergovernmental science-policy platform on biodiversity and ecosystem services, Intergovernmental Science-Policy Platform on Biodiversity and Ecosystem Services (IPBES), 2019.

[3] United Nations, United Nations Decade of Ocean Science for Sustainable Development (2021–2030). A/RES/72/73, 2017. https://www.un.org/en/development/desa/population/migration/generalassembly/docs/globalcompact/A_RES_72_73.pdf (accessed November 24, 2025).

[4] United Nations, United Nations Decade on Ecosystem Restoration (2021-2030). A/RES/73/284, 2019. https://docs.un.org/en/a/res/73/284 (accessed November 24, 2025).

[5] European commission, EU Biodiversity Strategy for 2030: Bringing nature back into our lives, (2020). https://doi.org/10.2779/048.

[6] European Parliament and Council, REGULATION (EU) 2024/1991 OF THE EUROPEAN PARLIAMENT AND OF THE COUNCIL of 24 June 2024 on nature restoration and amending Regulation (EU) 2022/869, 2024. http://data.europa.eu/eli/reg/2024/1991/oj.

[7] E. McLeod, G.L. Chmura, S. Bouillon, R. Salm, M. Björk, C.M. Duarte, C.E. Lovelock, W.H. Schlesinger, B.R. Silliman, A blueprint for blue carbon: Toward an improved understanding of the role of vegetated coastal habitats in sequestering CO2, Front Ecol Environ 9 (2011) 552–560. https://doi.org/10.1890/110004.

[8] L. Pendleton, D.C. Donato, B.C. Murray, S. Crooks, W.A. Jenkins, S. Sifleet, C. Craft, J.W. Fourqurean, J.B. Kauffman, N. Marbà, P. Megonigal, E. Pidgeon, D. Herr, D. Gordon, A. Baldera, Estimating Global “Blue Carbon” Emissions from Conversion and Degradation of Vegetated Coastal Ecosystems, PLoS One 7 (2012). https://doi.org/10.1371/journal.pone.0043542.

[9] The Council of the European Union, Council Directive 92/43/EEC on the conservation of natural habitats and of wild fauna and flora, (1992).

[10] European Parliament and the Council, Directive 2000/60/EC — framework for Community action in the field of water policy, (2000).

[11] European Parliament and the Council, Marine Strategy Framework Directive 2008/56/EC, 2008.

[12] J. Maes, B. Egoh, L. Willemen, C. Liquete, P. Vihervaara, J.P. Schägner, B. Grizzetti, E.G. Drakou, A. La Notte, G. Zulian, F. Bouraoui, M. Luisa Paracchini, L. Braat, G. Bidoglio, Mapping ecosystem services for policy support and decision making in the European Union, Ecosyst Serv 1 (2012) 31–39. https://doi.org/10.1016/j.ecoser.2012.06.004.

[13] A. Newton, J. Icely, S. Cristina, G.M.E. Perillo, R.E. Turner, D. Ashan, S. Cragg, Y. Luo, C. Tu, Y. Li, H. Zhang, R. Ramesh, D.L. Forbes, C. Solidoro, B. Béjaoui, S. Gao, R. Pastres, H. Kelsey, D. Taillie, N. Nhan, A.C. Brito, R. de Lima, C. Kuenzer, Anthropogenic, Direct Pressures on Coastal Wetlands, Front Ecol Evol 8 (2020). https://doi.org/10.3389/fevo.2020.00144.

[14] F. Borgwardt, L. Robinson, D. Trauner, H. Teixeira, A.J.A. Nogueira, A.I. Lillebø, G. Piet, M. Kuemmerlen, T. O'Higgins, H. McDonald, J. Arevalo-Torres, A.L. Barbosa, A. Iglesias-Campos, T. Hein, F. Culhane, Exploring variability in environmental impact risk from human activities across aquatic ecosystems, Science of the Total Environment 652 (2019) 1396–1408. https://doi.org/10.1016/j.scitotenv.2018.10.339.

[15] IPBES, Opportunities for communities of practice to engage with and contribute to the work of the Intergovernmental Platform on Biodiversity and Ecosystem Services (IPBES), 2021. https://ipbes.net/documents-by-category/policies.

[16] E. Kampa, E. Elkina, B. Bueb, M. del M. Otero Villanueva, Restoring European Coastal Wetlands for Climate and Biodiversity: Do EU Policies and International Agreements Support Restoration?, Sustainability (Switzerland) 17 (2025). https://doi.org/10.3390/su17219469.

[17] A.I. Sousa, J.F. da Silva, A. Azevedo, A.I. Lillebø, Blue Carbon stock in Zostera noltei meadows at Ria de Aveiro coastal lagoon (Portugal) over a decade, Sci Rep 9 (2019). https://doi.org/10.1038/s41598-019-50425-4.

[18] A.I. Lillebø, P. Stålnacke, G.D. Gooch, Coastal Lagoons in Europe - Integrated water resources strategies, 2015.

[19] V. Costa, M.R. Flindt, M. Lopes, J.P. Coelho, A.F. Costa, A.I. Lillebø, A.I. Sousa, Enhancing the resilience of Zostera noltei seagrass meadows against Arenicola spp. bio-invasion: A decision-making approach, J Environ Manage 302 (2022). https://doi.org/10.1016/j.jenvman.2021.113969.

[20] D. Crespo, R. Faião, V. Freitas, V.H. Oliveira, A.I. Sousa, J.P. Coelho, M. Dolbeth, Using seagrass as a nature-based solution: Short-term effects of Zostera noltei transplant in benthic communities of a European Atlantic coastal lagoon, Mar Pollut Bull 197 (2023). https://doi.org/10.1016/j.marpolbul.2023.115762.

[21] C. Rochera, A. Picazo, D. Morant, J. Miralles-Lorenzo, V. Sánchez-Ortega, A. Camacho, Linking Carbon Fluxes to Flooding Gradients in Sediments of Mediterranean Wetlands, ACS ES and T Water 5 (2025) 2882–2890. https://doi.org/10.1021/acsestwater.4c00940.

[22] A. Davranche, C. Arzel, P. Pouzet, A.R. Carrasco, G. Lefebvre, D. Lague, M. Thibault, A. Newton, C. Fleurant, M. Maanan, B. Poulin, A multi-sensor approach to monitor the ongoing restoration of edaphic conditions for salt marsh species facing sea level rise: An adaptive management case study in Camargue, France, Science of the Total Environment 908 (2024). https://doi.org/10.1016/j.scitotenv.2023.168289.

[23] P.W.J.M. Willemsen, E.M. Horstman, T.J. Bouma, M.J. Baptist, M.E.B. van Puijenbroek, B.W. Borsje, Facilitating Salt Marsh Restoration: The Importance of Event-Based Bed Level Dynamics and Seasonal Trends in Bed Level Change, Front Mar Sci 8 (2022). https://doi.org/10.3389/fmars.2021.793235.

[24] I. Vybernaite-Lubiene, M. Zilius, L. Saltyte-Vaisiauske, M. Bartoli, Recent Trends (2012-2016) of N, Si, and P export from the Nemunas River Watershed: Loads, unbalanced stoichiometry, and threats for downstream aquatic ecosystems, Water (Switzerland) 10 (2018). https://doi.org/10.3390/w10091178.

[25] M. Bartoli, M. Zilius, M. Bresciani, D. Vaiciute, I. Vybernaite-Lubiene, J. Petkuviene, G. Giordani, D. Daunys, T. Ruginis, S. Benelli, C. Giardino, P.A. Bukaveckas, P. Zemlys, E. Griniene, Z.R. Gasiunaite, J. Lesutiene, R. Pilkaityte, A. Baziukas-Razinkovas, Drivers of cyanobacterial blooms in a hypertrophic lagoon, Front Mar Sci 5 (2018). https://doi.org/10.3389/fmars.2018.00434.

[26] J. Petkuviene, M. Zilius, I. Lubiene, T. Ruginis, G. Giordani, A. Razinkovas-Baziukas, M. Bartoli, Phosphorus Cycling in a Freshwater Estuary Impacted by Cyanobacterial Blooms, Estuaries and Coasts 39 (2016) 1386–1402. https://doi.org/10.1007/s12237-016-0078-0.

[27] M. Zilius, M. Bartoli, M. Bresciani, M. Katarzyte, T. Ruginis, J. Petkuviene, I. Lubiene, C. Giardino, P.A. Bukaveckas, R. de Wit, A. Razinkovas-Baziukas, Feedback mechanisms between cyanobacterial blooms, transient hypoxia, and benthic phosphorus regeneration in shallow coastal environments, Estuaries and Coasts 37 (2014) 680–694. https://doi.org/10.1007/s12237-013-9717-x.

[28] Z. Sinkevičiene, M. Bučas, R. Ilgine, Di. Vaičiute, M. Katarzyte, J. Petkuviene, Charophytes in the estuarine Curonian Lagoon: Have the changes in diversity, abundance and distribution occurred since the late 1940s?, Oceanol Hydrobiol Stud 46 (2017) 186–198. https://doi.org/10.1515/ohs-2017-0019.

[29] G. Lupu, S. Covaliov, M. Doroftei, M. Mierlă, M. Simionov, A. Năstase, I. Cenișencu, I. Fomici, S.-D. Chirilă, A. Naum, A. Doroșencu, L. Bolboacă, M. Marinov, V. Alexe, F. Sicrieru, L. Ene, Status of biodiversity, reed habitats, sustainable exploitation of natural resources, invasive species, and socio-economic implications in Danube Delta Biosphere Reserve in 2024, Scientific Annals of the Danube Delta Institute 30 (2025) 141–162. https://doi.org/10.3897/saddi.30.163613.

[30] K.A. Aivaz, M. Serbanescu, Ecosystem Services Evaluation of the Danube Delta: An Analysis Using Hierarchical Multifactor Regression, Studies in Business and Economics 19 (2024) 5–21. https://doi.org/10.2478/sbe-2024-0001.

[31] R. Fradoca, V.H. Oliveira, B.A. Fonte, A.I. Sousa, B. Marques, A.I. Lillebø, J.P. Coelho, The effect of Zostera noltei recolonization on N and P fluxes at the sediment/water interface, Mar Pollut Bull 215 (2025). https://doi.org/10.1016/j.marpolbul.2025.117901.

[32] G. Umgiesser, P. Zemlys, A. Erturk, A. Razinkova-Baziukas, J. Mezine, C. Ferrarin, Seasonal renewal time variability in the Curonian Lagoon caused by atmospheric and hydrographical forcing, Ocean Science 12 (2016) 391–402. https://doi.org/10.5194/os-12-391-2016.

[33] D. Vaičiūtė, M. Bučas, M. Bresciani, T. Dabulevičienė, J. Gintauskas, J. Mėžinė, E. Tiškus, G. Umgiesser, J. Morkūnas, F. De Santi, M. Bartoli, Hot moments and hotspots of cyanobacteria hyperblooms in the Curonian Lagoon (SE Baltic Sea) revealed via remote sensing-based retrospective analysis, Science of the Total Environment 769 (2021). https://doi.org/10.1016/j.scitotenv.2021.145053.

[34] F. Christian, R. Arturas, G. Saulius, U. Georg, B. Lina, Hydraulic regime-based zonation scheme of the Curonian Lagoon, Hydrobiologia 611 (2008) 133–146. https://doi.org/10.1007/s10750-008-9454-5.

[35] E. Trimonis, S. Gulbinskas, M. Kuzavinis, The Curonian Lagoon bottom sediments in the Lithuanian water area, Baltica (2003) 1320. http://scholar.google.com/scholar_lookup?&title=The+Curonian+Lagoon+bottom+sediments+in+the+Lithuanian+water+area%2E&journal=Baltica&author=Trimonis+E.&author=Gulbinskas+S.&author=and+Kuzavinis+M.&publication_year=2003&volume=16&pages=13-20. (accessed November 24, 2025).

[36] T. Politi, M. Zilius, P. Forni, A. Zaiko, D. Daunys, M. Bartoli, Biogeochemical buffers in a eutrophic coastal lagoon along an oxic-hypoxic transition, Estuar Coast Shelf Sci 279 (2022). https://doi.org/10.1016/j.ecss.2022.108132.

[37] European Environment Agency (EEA), Ecosystem types of Europe 2012 - Full map (marine and terrestrial habitats) - version 3 revision [Dataset], European Environment Agency (2019). https://www.eea.europa.eu/en/datahub/datahubitem-view/573ff9d5-6889-407f-b3fc-cfe3f9e23941?activeAccordion=1069819%2C1069817%2C1069818%2C1069816 (accessed September 15, 2025).

[38] European Environment Agency (EEA), EUNIS terrestrial habitat classification review (tabular) - version 1, Nov. 2021, European Environment Agency (EEA) (2021). https://doi.org/https://doi.org/10.2909/bfe4c237-e378-4a83-ab21-b3807f96c2e2.

[39] European Environment Agency (EEA), EUNIS marine habitat classification review (tabular) - version 1, 2022, European Environment Agency (EEA) (2022). https://doi.org/https://doi.org/10.2909/8a5eccda-e6ea-4018-a373-5c76a8eeec78.

[40] R. Haines-Young, Common International Classification of Ecosystem Services (CICES) V5.2 Guidance on the Application of the Revised Structure, 2023. www.cices.eu].

[41] M. Chytrý, L. Tichý, S.M. Hennekens, I. Knollová, J.A.M. Janssen, J.S. Rodwell, T. Peterka, C. Marcenò, F. Landucci, J. Danihelka, M. Hájek, J. Dengler, P. Novák, D. Zukal, B. Jiménez-Alfaro, L. Mucina, S. Abdulhak, S. Aćić, E. Agrillo, F. Attorre, E. Bergmeier, I. Biurrun, S. Boch, J. Bölöni, G. Bonari, T. Braslavskaya, H. Bruelheide, J.A. Campos, A.

Čarni, L. Casella, M. Ćuk, R. Ćušterevska, E. De Bie, P. Delbosc, O. Demina, Y. Didukh, D. Dítě, T. Dziuba, J. Ewald, R.G. Gavilán, J.C. Gégout, G. Pietro Giusso del Galdo, V. Golub, N. Goncharova, F. Goral, U. Graf, A. Indreica, M. Isermann, U. Jandt, F. Jansen, J. Jansen, A. Jašková, M. Jiroušek, Z. Kącki, V. Kalníková, A. Kavgacı, L. Khanina, A. Yu. Korolyuk, M. Kozhevnikova, A. Kuzemko, F. Küzmič, O.L. Kuznetsov, M. Laiviņš, I. Lavrinenko, O. Lavrinenko, M. Lebedeva, Z. Lososová, T. Lysenko, L. Maciejewski, C. Mardari, A. Marinšek, M.G. Napreenko, V. Onyshchenko, A. Pérez-Haase, R. Pielech, V. Prokhorov, V. Rašomavičius, M.P. Rodríguez Rojo, S. Rūsiņa, J. Schrautzer, J. Šibík, U. Šilc, Ž. Škvorc, V.A. Smagin, Z. Stančić, A. Stanisci, E. Tikhonova, T. Tonteri, D. Uogintas, M. Valachovič, K. Vassilev, D. Vynokurov, W. Willner, S. Yamalov, D. Evans, M. Palitzsch Lund, R. Spyropoulou, E. Tryfon, J.H.J. Schaminée, EUNIS Habitat Classification: Expert system, characteristic species combinations and distribution maps of European habitats, Appl Veg Sci 23 (2020) 648–675. https://doi.org/10.1111/avsc.12519.

[42] Bruna R.F. Oliveira, António Nogueira, Ana Lillebø, Report on the assessment of co-benefits and economic valuation of ecosystem services provision - RESTORE4Cs Deliverable 5.2, 2025. https://www.restore4cs.eu/ (accessed December 2, 2025).

[43] Ł. Janowski, A. Barańska, K. Załęski, M. Kubacka, M. Michałek, A. Tarała, M. Niemkiewicz, J. Gajewski, Predictive Benthic Habitat Mapping Reveals Significant Loss of Zostera marina in the Puck Lagoon, Baltic Sea, over Six Decades, Remote Sens (Basel) 17 (2025) 3725. https://doi.org/10.3390/rs17223725.

[44] S. Si-Moussi, S. Hennekens, S. Mücher, W. De Keersmaecker, M. Chytrý, E. Agrillo, F. Attorre, I. Biurrun, G. Bonari, A. Čarni, R. Ćušterevska, T. Dziuba, K. Ecker, B. Güler, U. Jandt, B. Jiménez-Alfaro, J. Lenoir, J.-C. Svenning, G. Swacha, W. Thuiller, EUNIS Habitat Maps: Enhancing Thematic and Spatial Resolution for Europe through Machine Learning, 2025.

[45] B. Grizzetti, C. Liquete, A. Pistocchi, O. Vigiak, G. Zulian, F. Bouraoui, A. De Roo, A.C. Cardoso, Relationship between ecological condition and ecosystem services in European rivers, lakes and coastal waters, Science of the Total Environment 671 (2019) 452–465. https://doi.org/10.1016/j.scitotenv.2019.03.155.

[46] H. Teixeira, A.I. Lillebø, F. Culhane, L. Robinson, D. Trauner, F. Borgwardt, M. Kummerlen, A. Barbosa, H. McDonald, A. Funk, T. O'Higgins, J.T. Van der Wal, G. Piet, T. Hein, J. Arévalo-Torres, A. Iglesias-Campos, J. Barbière, A.J.A. Nogueira, Linking biodiversity to ecosystem services supply: Patterns across aquatic ecosystems, Science of the Total Environment 657 (2019) 517–534. https://doi.org/10.1016/j.scitotenv.2018.11.440.

[47] Timothy G. O'Higgins, Manuel Lago, Theodore H. DeWitt, Ecosystem-Based Management, Ecosystem Services and Aquatic Biodiversity, Springer International Publishing, Cham, 2020. https://doi.org/10.1007/978-3-030-45843-0.

[48] Lisa Sella, Francesca Silvia Rota, Nicola Pollo, Gianna Vivaldo, Clémentina Anglada, Giulia de Fusco, Elisa Ciravegna, Social acceptability of wetland restoration and

management - RESTORE4Cs Deliverable 5.4, 2025. https://www.restore4cs.eu/ (accessed December 2, 2025).

[49] Clémentine Anglada, Jade Massoutier, Manuel Lago, Elisa Ciravegna, Justine Raoult, Nico Polman, Auriane Bodivit, Lisa Sella, Report on cost/benefit analysis of wetland restoration options and on financing tools - RESTORE4Cs Deliverable 5.3, 2025. https://www.restore4cs.eu/ (accessed December 2, 2025).

[50] M. Marttunen, J. Mustajoki, M. Dufva, TimoP. Karjalainen, How to design and realize participation of stakeholders in MCDA processes? A framework for selecting an appropriate approach, EURO Journal on Decision Processes 3 (2015) 187–214. https://doi.org/10.1007/s40070-013-0016-3.